\documentclass[10pt,final,twocolumn,conference]{IEEEtran}
\usepackage{graphicx}

\usepackage{amsmath}
\usepackage{amssymb}
\usepackage{fancyhdr}
\usepackage{multicol}
\usepackage{multirow}
\usepackage{algorithm}
\usepackage[noend]{algpseudocode}
\usepackage{eso-pic}
\usepackage{color}
\usepackage{tikz}
\usetikzlibrary{shapes}

\begin{document}
\newcommand{\unfilledsquare}{\raisebox{0.5pt}{\tikz{\node[draw,scale=0.4,regular polygon, regular polygon sides=4,fill=none](){};}}}
\newcommand{\bluediamond}{\raisebox{0pt}{\tikz{\node[fill,scale=0.4,diamond,fill=black!10!blue](){};}}}
\newcommand{\greenbox}{\raisebox{0.5pt}{\tikz{\node[draw=green!80, thick, scale=0.4, regular polygon, regular polygon sides=4,fill=none](){};}}}
\newcommand{\reddottedbox}{\raisebox{0.5pt}{\tikz{\node[draw=red!80, thick, scale=0.4, regular polygon, regular polygon sides=4,fill=none](){};}}}
\renewcommand{\vec}[1]{\boldsymbol{#1}}

\author{
	\IEEEauthorblockN{Yeong Foong Choo\IEEEauthorrefmark{1}, Brian L. Evans\IEEEauthorrefmark{1} and Alan Gatherer\IEEEauthorrefmark{2}\\}
	\IEEEauthorblockA{\IEEEauthorrefmark{1}Wireless Networking and Communications Group, The University of Texas at Austin, Austin, TX USA}
	\IEEEauthorblockA{\IEEEauthorrefmark{2}Wireless Access Laboratory, Huawei Technologies, Plano, TX USA}
	\IEEEauthorrefmark{1}{yeongfoong.choo@utexas.edu, bevans@ece.utexas.edu}
	\IEEEauthorrefmark{2}{alan.gatherer@huawei.com}
}
\title{\huge Complex Block Floating-Point Format with Box Encoding For Wordlength Reduction in Communication Systems}

\maketitle

\begin{abstract}
We propose a new complex block floating-point format to reduce implementation complexity.
The new format achieves wordlength reduction by sharing an exponent across the block of samples, and uses box encoding for the shared exponent to reduce quantization error. Arithmetic operations are performed on blocks of samples at time, which can also reduce implementation complexity. For a case study of a baseband quadrature amplitude modulation (QAM) transmitter and receiver, we quantify the tradeoffs in signal quality vs. implementation complexity using the new approach to represent IQ samples. Signal quality is measured using error vector magnitude (EVM) in the receiver, and implementation complexity is measured in terms of arithmetic complexity as well as memory allocation and memory input/output rates. The primary contributions of this paper are (1) a complex block floating-point format with box encoding of the shared exponent to reduce quantization error, (2) arithmetic operations using the new complex block floating-point format, and (3) a QAM transceiver case study to quantify signal quality vs. implementation complexity tradeoffs using the new format and arithmetic operations.

\end{abstract}


\section{Introduction}

Energy-efficient data representation in application specific baseband transceiver hardware are in demand resulting from energy costs involved in baseband signal processing \cite{fettweis2008ict}. In macrocell base stations, about ten percent of energy cost contribute towards digital signal processing (DSP) modules while power amplification and cooling processes consume more than 70\% of total energy \cite{5463328}. The energy consumption by DSP modules relative to power amplification and cooling will increase in future designs of small cell systems because low-powered cellular radio access nodes handle a shorter radio range \cite{5463328}. The design of energy-efficient number representation will reduce overall energy consumption in base stations.

In similar paper, baseband signal compression techniques have been researched for both uplink and downlink. The methods in \cite{6226311}, \cite{6737122}, and \cite{7054343} suggest resampling baseband signals to Nyquist rate, block scaling, and non-linear quantization. All three papers report transport data rate gain of 3x to 5x with less than 2\% EVM loss. In \cite{7054343}, cyclic prefix replacement technique is used to counter the effect of resampling, which would add processing overhead to the system. In \cite{6737122} and \cite{7094798}, noise shaping technique shows improvement of in-band signal-to-noise ratio (SNR). In \cite{7511265}, transform coding technique is suggested for block compression of baseband signals in the settings of multiple users and multi-antenna base station. Transform coding technique reports potential of 8x transport data rate gain with less than 3\% EVM loss. The above methods achieve end-to-end compression in a transport link and incur delay and energy cost for the compression and decompression at the entry and exit points, respectively. The overall energy cost reduction is not well quantified. This motivates the design of energy-efficient data representation and hardware arithmetic units with low implementation complexity.

In \cite{6169957}, Common Exponent Encoding is proposed to represent 32-bit complex floating-point data by only 29-bit wordlength in hardware to achieve 3-bit savings. The method in \cite{6169957} shows 10\% reduction of registers and memory footprints with a tradeoff of 10\% increase in arithmetic units. In \cite{7178126}, exponential coefficient scaling is proposed to allocate 6 bits to represent real-valued floating-point data. The method in \cite{7178126} achieves 37x reduction in quantization errors, 1.2x reduction in logic gates, and 1.4x reduction in energy per cycle compared to 6-bit fixed-point representation. Both papers report less than 2 dB of signal-to-quantization-noise ratio (SQNR).

\begin{figure}
	\includegraphics[width=3.5in]{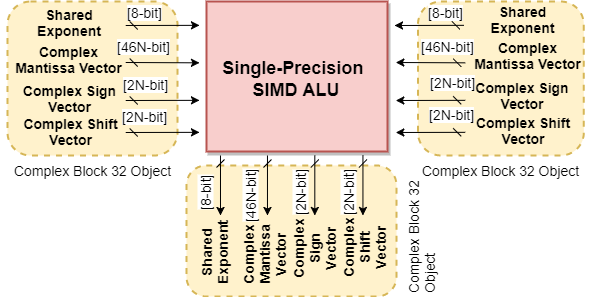}
	\caption{ 32-bit equivalent SIMD ALU in Exponent Box Encoding format }
	\label{fig:32bitSIMD_ALU}
\end{figure}

\textbf{\textit{Contributions:}} Our method applies the Common Exponent Encoding proposed by \cite{6169957} and adds a proposed Exponent Box Encoding to retain high magnitude-phase resolution. This paper identifies the computational complexity of complex block addition, multiplication, and convolution and computes reference EVM on the arithmetic output. We apply the new complex block floating-point format to case study of baseband QAM transmitter chain and receiver chain. We also reduce implementation complexity in terms of memory reads/writes rates, and multiply-accumulate operations. We base the signal quality of our method on the measurement of EVM at the receiver. Our method achieves end-to-end complex block floating-point representation.

\section{Methods}
\label{methodsSection}
This section describes the data structure used in new representation of complex block floating-point \cite{6169957} and suggests a new mantissa scaling method in reducing quantization error. In IEEE 754 format, the exponents of complex-valued floating-point data are separately encoded. Common Exponent Encoding technique \cite{6169957} allows common exponent sharing that has weak encoding of phase resolution.

\subsection{Common Exponent Encoding Technique}
Table \ref{tab:IEEEFloatingPointFormat} summarizes the wordlength precision of real-valued floating-point data in IEEE-754 encoding \cite{4610935}. We define \(B_w\)-bit as the wordlength of scalar floating-point data. A complex-valued floating-point data requires \(2B_w\)-bit and a complex block floating-point of \(N_v\) samples requires \(2N_vB_w\)-bit.

\begin{table}
	\centering
	\caption{Definition \& Bit Widths Under IEEE-754 Number Format \cite{4610935}}
	\label{tab:IEEEFloatingPointFormat}

		\begin{tabular}{| c || c | c |}
		\hline
		  \textbf{Components} & \textbf{Definition} & \textbf{Bit Widths, \(B\)} \\ \hline \hline
			Wordlength, \(W\)   & \(N_w\)    & \(\{16, 32, 64\}\) \\ \hline 
			Sign, \(S\)         & \(N_s\)    & \(\{1\}\) \\ \hline 
			Exponent, \(E\)     & \(N_e\)    & \(\{5, 8, 11\}\) \\ \hline 
			Mantissa, \(M\)     & \(N_m\)    & \(\{10, 23, 52\}\) \\ \hline 
		\end{tabular}
\end{table}

\begin{table}
	\centering
	\caption{Definition \& Bit Widths Under Common Exponent Encoding \cite{6169957}}
	\label{tab:CommonExponentFormat}

		\begin{tabular}{| c || c | c |}
		\hline
		  \textbf{Components}           & \textbf{Definition}   & \textbf{Bit Widths, \(B\)} \\ \hline \hline
			Common Exponent, \(E\)        & \(N_{e}\)             & \(\{5, 8, 11\}\) \\ \hline 
			Real / Imaginary, \(\vec{S}\)       & \(N_{s}^{R,I}\)       & \(\{1\}\) \\ \hline 
			Real / Imaginary Lead, \(\vec{L}\)  & \(N_{l}^{R,I}\)       & \(\{1\}\) \\ \hline 
			Real / Imaginary Mantissa, \(\vec{M}\)  & \(N_{m}^{R,I}\)   & \(\{10, 23, 52\}\) \\ \hline 
		\end{tabular}
\end{table}

\begin{figure}
	\includegraphics[width=3.5in]{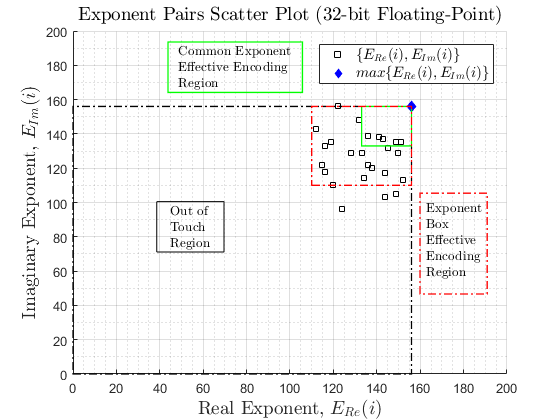}
	\caption{Scatter plot of $N = 25$ complex-valued exponent pairs $X \sim N(130,12^2) $ (\protect\unfilledsquare) and potential candidate for common exponent (\protect\bluediamond) }
	\label{fig:effectiveExponentEncoding}
\end{figure}

The method in \cite{mcgowan2015compact} assumes only magnitude correlation in the oversampled complex block floating-point data. This assumption allows common exponent be jointly encoded across complex block floating-point of \(N_v\) samples defined in Table \ref{tab:CommonExponentFormat}. The implied leading bit of 1 of each floating-point data is first uncovered. The common exponent is selected from the largest unsigned exponent across the complex block. All mantissa values are successively scaled down by the difference between common exponent and its original exponent. Therefore, each floating-point data with smaller exponents value loses leading bit of 1. The leading bit of complex block floating-point is explicitly coded as \(N_l\), using \(B_l\)-bit. The sign bits are left unchanged. A complex block floating-point of \(N_v\) samples requires \(\{2N_v(B_s + B_l + B_m) + B_e\}\)-bit.

We derive the maximum allowed exponent difference under Common Exponent Encoding in Appendix \ref{DerivationMaximumExponentDifferenceAppendix}. Mantissa values could be reduced to zero as a result of large phase difference. Figure \ref{fig:effectiveExponentEncoding} shows the Effective Encoding Region (EER) under Common Exponent Encoding technique (\protect\greenbox). Exponent pairs outside the EER will have corresponding mantissa values reduce to zero.

\subsection{Exponent Box Encoding Technique}
\begin{table}
	\centering
	\caption{Definition \& Bit Widths Under Exponent Box Encoding}
	\label{tab:ExponentBoxShiftFormat}

		\begin{tabular}{| c || c | c |}
		\hline
		  \textbf{Components} & \textbf{Definition} & \textbf{Bit Widths , \(B\)} \\ \hline \hline
			Common Exponent, \(E\)             & \(N_{e}\)         & \(\{5, 8, 11\}\) \\ \hline 
			Real / Imaginary Sign, \(\vec{S}\)       & \(N_{s}^{R,I}\)   & \(\{1\}\) \\ \hline 
			Real / Imaginary Lead, \(\vec{L}\)       & \(N_{l}^{R,I}\)   & \(\{1\}\) \\ \hline 
			Real / Imaginary Box Shift, \(\vec{X}\)  & \(N_{x}^{R,I}\)   & \(\{1\}\) \\ \hline 
			Real / Imaginary Mantissa, \(\vec{M}\)   & \(N_{m}^{R,I}\)   & \(\{10, 23, 52\}\) \\ \hline 
		\end{tabular}
\end{table}

The Common Exponent Encoding technique suffers high quantization and phase error in the complex block floating-point of high dynamic range. Exponent Box Encoding is suggested to reduce quantization error of complex-valued floating-point pairs by allocating \(2N_v\)-bit per complex block. Figure \ref{fig:effectiveExponentEncoding} shows the Effective Encoding Region under Exponent Box Encoding technique (\protect\reddottedbox) which has four times larger the area of EER of Common Exponent Encoding technique (\protect\greenbox).

The use of 2-bit per complex sample replaces the mantissas rescaling operation with exponents addition/ subtraction. We are able to preserve more leading bits of mantissas values which improve the accuracy of complex block multiplication and complex block convolution results. A complex block floating-point of \(N_v\) samples requires \(\{2N_v(B_s + B_l + B_x + B_m) + B_e\}\)-bit.

Arithmetic Logic Unit (ALU) hardware is designed to perform Single-Instruction Multiple-Data (SIMD) operation on complex block floating-point data. The Exponent Box Encoding is performed when converting to Exponent Box Encoding format. The Exponent Box Decoding is performed at the pre-processing of mantissas in Complex Block Addition and pre-processing of exponents in Complex Block Multiply.

Table \ref{tab:wordlengthrequirement} summarizes the wordlength analysis required by complex block floating-point of \(B_v\) samples. The Exponent Box Encoding and Exponent Box Decoding algorithms are described as follows:

\begin{algorithm}
\caption{Exponent Box Encoding}
\label{alg:exponentBoxEncode}
\begin{algorithmic}
\State $Let \quad U \gets {max\{E\{s\}\} - B_m} $
\For{ $ i^{th} \in N_v \quad \{R/I\}\,samples $}
\If { $ N_e^{R}\{i\} < U$}
\State $ N_e^{R}\{i\} \gets N_e^{R}\{i\} + B_m $ 
\State $ N_x^{R}\{i\} \gets 1 $
\EndIf
\If { $ N_e^{I}\{i\} < U$}
\State $ N_e^{I}\{i\} \gets N_e^{I}\{i\} + B_m $ 
\State $ N_x^{I}\{i\} \gets 1 $
\EndIf
\EndFor
\end{algorithmic}
\end{algorithm}

\begin{algorithm}
\caption{Exponent Box Decoding}
\label{alg:exponentBoxDecode}
\begin{algorithmic}
\For{ $ i^{th} \in N_v \quad \{R/I\}\,samples $ }
\If { $ N_x^{R}\{i\} \equiv 1 $ }
\State $ N_e^{R}\{i\} \gets N_e^{R}\{i\} - B_m $ 
\EndIf
\If { $ N_x^{I}\{i\} \equiv 1 $}
\State $ N_e^{I}\{i\} \gets N_e^{I}\{i\} - B_m $ 
\EndIf
\EndFor
\end{algorithmic}
\end{algorithm}

\begin{table}
	\centering
	\caption{Wordlength Requirement by \(N_v\) Complex-Valued Samples}
	\label{tab:wordlengthrequirement}

		\begin{tabular}{| c || c |}
		\hline
			\textbf{Encoding} & \textbf{Bit Widths}            \\ \hline \hline
			Complex IEEE754 & \(2N_v(B_s+B_e+B_m)\)          \\ \hline
			Common Exponent & \(2N_v(B_s+B_l+B_m)+B_e\)      \\ \hline
			Exponent Box    & \(2N_v(B_s+B_l+B_x+B_m)+B_e\)  \\ \hline
		\end{tabular}
\end{table}

\begin{figure}
	\includegraphics[width=3.5in]{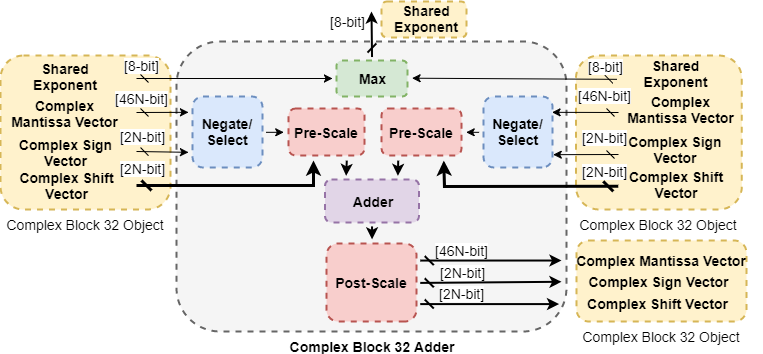}
	\caption{Pre-Scale mantissas in Complex Block Add}
	\label{fig:SIMD_Add}
\end{figure}

\section{Arithmetic Unit}
\label{arithmeticSection}

We identify the arithmetic units predominantly used on complex block floating-point data. Complex-valued multiplication and addition are two primary ALU required in convolution operation. This section identifies the complexity of pre-processing and post-processing mantissas and exponents in the complex block addition, multiplication, and convolution arithmetic. Table \ref{tab:exponentMantissasRequirement} describes the worst-case complexity analysis of complex block ALU on encoding format described in Section \ref{methodsSection}.

\begin{table*}[!t]
	\centering
	\caption{Mantissas and Exponent Pre/Post Processing Complexity of Complex Block ALU}
	\label{tab:exponentMantissasRequirement}

		\begin{tabular}{| c || c | c |}
		\hline
			\textbf{Block Addition} & \textbf{Mantissas Scaling} & \textbf{Exponents Arithmetic}  \\ \hline \hline
			Complex IEEE754       & \(4*N\)         & \(2*N\)    \\ \hline 
			Common Exponent       & \(4*N\)         & \(2\)      \\ \hline 
			Exponent Box          & \(8*N\)         & \(4\)      \\ \hline \hline
			
			\textbf{Block Multiplication} & \textbf{Mantissas Scaling} & \textbf{Exponents Arithmetic}  \\ \hline \hline
			Complex IEEE754       & \(8*N\)         & \(6*N\)    \\ \hline 
			Common Exponent       & \(8*N\)         & \(2\)      \\ \hline 
			Exponent Box          & \(16*N\)        & \(5\)      \\ \hline \hline
			
			\textbf{Convolution} & \textbf{Mantissas Scaling} & \textbf{Exponents Arithmetic} \\ \hline \hline
			Complex IEEE754			  & \(6*N_1N_2+4*(N_1-1)(N_2-1)\) & \(6*N_1N_2+2*(N_1-1)(N_2-1)\) \\ \hline 
			Common Exponent       & \(6*N_1N_2+4*(N_1-1)(N_2-1)\) & \(3*(N_1+N_2-1) + 1\) \\ \hline 
			Exponent Box          & \(10*N_1N_2+8*(N_1-1)(N_2-1)\)& \(3*(N_1+N_2-1) + 1\) \\ \hline \hline
			
		\end{tabular}
\end{table*}

\subsection{Complex Block Addition}
Figure \ref{fig:SIMD_Add} shows simplified block diagram for Complex Block Addition. 
Let $\vec{X_1,X_2,Y}$ \( \in \mathbb{C}^{1 \times N}\) be complex-valued row vectors, such that,

\begin{equation} \label{eq4}
\begin{split}
\Re \{ \vec{Y} \} &= \Re \{ \vec{X_1} \} + \Re \{ \vec{X_2} \} \\
\Im \{ \vec{Y} \} &= \Im \{ \vec{X_1} \} + \Im \{ \vec{X_2} \} \\
\end{split}
\end{equation}

In IEEE-754 encoding format, complex block addition is implemented as two real-valued addition. There are four exponents to the two complex inputs and two exponents to the complex output. Each real-valued addition block requires one mantissa pre-scaling, one mantissa post-scaling, and one exponent arithmetic. Therefore, complex block addition requires two mantissas pre-scaling, two mantissas post-scaling, and two exponents arithmetic per sample.

In Common Exponent and Exponent Box Encoding, there are two shared exponents to the two complex block inputs and one shared exponent to the complex block output. Complexity on shared exponent arithmetic is \(O(1)\). We pre-scale the mantissas corresponding to the smaller exponent and post-scale the mantissas of the complex block output. With Exponent Box Encoding in the worst case, we require two mantissas pre-scaling and one mantissas post-scaling.

\begin{figure}
	\includegraphics[width=3.5in]{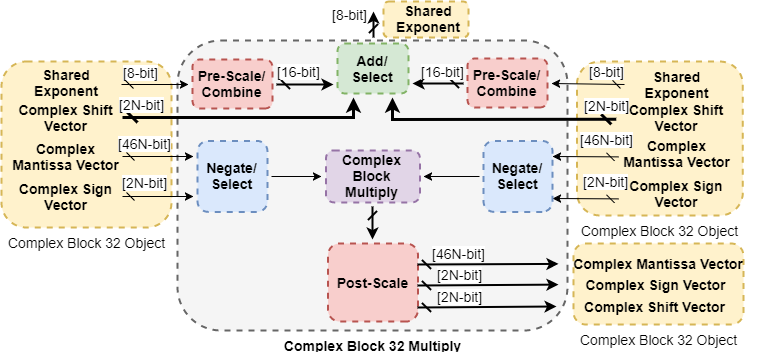}
	\caption{Pre-Scale exponents in Complex Block Multiply}
	\label{fig:SIMD_Multiply}
\end{figure}

\subsection{Complex Block Multiplication}
Figure \ref{fig:SIMD_Multiply} shows simplified block diagram for Complex Block Multiplication. Let $\vec{X_1,X_2,Y}$ \( \in \mathbb{C}^{1 \times N}\) be complex-valued row vectors, where \(\bullet\) denotes element-wise multiply, such that,

\begin{equation} \label{eq5}
\begin{split}
\Re \{ \vec{Y} \} &= \Re \{ \vec{X_1} \} \bullet \Re \{ \vec{X_2} \} - \Im \{ \vec{X_1} \} \bullet \Im \{ \vec{X_2} \} \\
\Im \{ \vec{Y} \} &= \Re \{ \vec{X_1} \} \bullet \Im \{ \vec{X_2} \} + \Im \{ \vec{X_1} \} \bullet \Re \{ \vec{X_2} \}\\
\end{split}
\end{equation}

In IEEE-754 encoding format, complex block multiplication is implemented as four real-valued multiplication and two real-valued addition. Each real-valued multiplication requires one mantissa post-scaling and one exponent arithmetic. Each real-valued addition requires one mantissa pre-scaling, one mantissa post-scaling, and one exponent arithmetic. Complex block multiply requires two mantissas pre-scaling, six mantissas post-scaling, and six exponent arithmetic per sample.

In Common Exponent and Exponent Box Encoding, we need two exponent arithmetic for multiply and normalization of the complex block output. With Exponent Box Encoding in the worst case, we need eight more mantissas post-scaling. Also, the Shift Vectors allow for four possible intermediate exponent values instead of one intermediate exponent value in Common Exponent Encoding.

\subsection{Complex Convolution}
Let $\vec{X_1}$ \( \in \mathbb{C}^{1 \times N_1}\), $\vec{X_2}$ \( \in \mathbb{C}^{1 \times N_2}\), and $\vec{Y}$ \( \in \mathbb{C}^{1 \times (N_1+N_2-1)}\) be complex-valued row vectors, where \(\ast\) denotes convolution, such that,

\begin{equation} \label{eq6}
\begin{split}
\Re \{ \vec{Y} \} &= {\Re \{ \vec{X_1} \ast \vec{X_2} \} }\\
\Im \{ \vec{Y} \} &= {\Im \{ \vec{X_1} \ast \vec{X_2} \} }\\
\end{split}
\end{equation}

We assume \(N_1 < N_2\) for practical reason where the model of channel impulse response has shorter sequence than the discrete-time samples. Each term in the complex block output is complex inner product of two complex block input of varying length between 1 and \(min\{N_1,N_2\}\). Complex convolution is implemented as complex block multiplication and accumulation of intermediate results. We derive the processing complexity of mantissas and exponents in Appendix \ref{ComplexConvolutionComplexityAppendix}.

\section{System Model}
\label{systemModel}
\begin{figure}
	\includegraphics[width=3.5in]{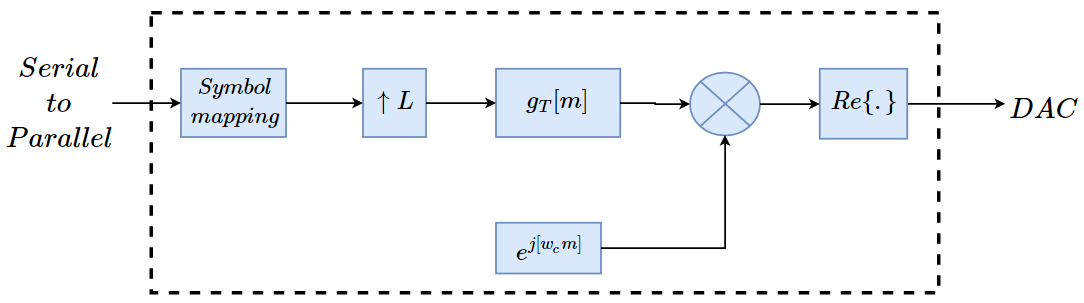}
	\caption{Block diagram of discrete-time complex baseband QAM transmitter}
	\label{fig:qamtransmit}
\end{figure}

\begin{figure}
	\includegraphics[width=3.5in]{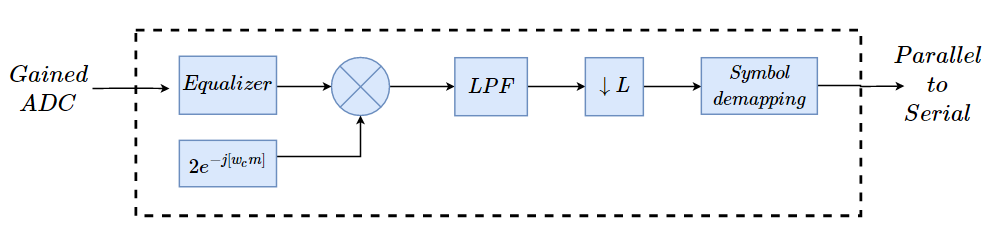}
	\caption{Block diagram of discrete-time complex baseband QAM receiver}
	\label{fig:qamreceive}
\end{figure}

We apply Exponent Box Encoding to represent IQ components in baseband QAM transmitter in Figure \ref{fig:qamtransmit} and baseband QAM receiver in Figure \ref{fig:qamreceive}. The simulated channel model is Additive White Gaussian Noise (AWGN). Table \ref{tab:qamTransceiverSpecs} contains the parameter definitions and values used in MATLAB simulation and Table \ref{tab:wordLengthRateAnalysisTransceiver} summarizes the memory input/output rates (bits/sec) and multiply-accumulate rates required by discrete-time complex QAM transmitter and receiver chains.

\begin{table}
	\centering
	\caption{QAM Transmitter, Receiver Specifications}
	\label{tab:qamTransceiverSpecs}

		\begin{tabular}{| c || c | c |}
		\hline
			QAM Parameters           & Definition                   & Values / Types        \\ \hline \hline			
			Constellation Order      & \(M\)                        & 1024          \\ \hline \hline
			
			Transceiver Parameters   & Definition                   & Values / Types      \\ \hline \hline			
			Up-sample Factor         & \(L^{TX},L^{RX}\)            & 4                \\ \hline
			Symbol Rate (Hz)         & \(f_{sym}\)                  & 2400               \\ \hline
			Filter Order             & \(N^{TX},N^{RX}\)            & \(32^{th}\)      \\ \hline
			Pulse Shape              & \(g^{TX},g^{RX}\)            & Root-Raised Cosine \\ \hline
			Excess Bandwidth Factor  & \(\alpha^{Tx},\alpha^{RX}\)  & 0.2               \\ \hline \hline
												
		\end{tabular}
\end{table}

\subsection{Discrete-time Complex Baseband QAM Transmitter}

\begin{table*}[!t]
	\centering
	\caption{Memory Input / Output and Computational Rates on Exponent Box Shifting Technique}
	\label{tab:wordLengthRateAnalysisTransceiver}

		\begin{tabular}{| c || c | c | c |}
		\hline
			Transmitter Chain & Memory Reads Rate (bits/sec) & Memory Writes Rate (bits/sec) & MACs / sec \\ \hline \hline
			Symbol Mapper      & \(Jf_{sym}\) & \(2f_{sym}(N_w+N_l+N_b-N_e)+N_e\) & \(0\) \\ \hline
			Upsampler          & \(2f_{sym}(N_w+N_l+N_b-N_e)+N_e\) & \(2L^{Tx}f_{sym}(N_w+N_l+N_b-N_e)+N_e\) & \(0\) \\ \hline
			Pulse Shape Filter & \((3L^{Tx}N_g^{Tx}+1)(L^{Tx}f_{sym})(N_w+N_l+N_b-N_e) + 2N_e\) & \(2L^{Tx}f_{sym}(N_w+N_l+N_b-N_e)+N_e\) & \((L^{Tx})^2N_g^{Tx}f_{sym}\) \\ \hline \hline
			Receiver Chain     & Memory Reads Rate (bits/sec) & Memory Writes Rate (bits/sec) & MACs / sec \\ \hline \hline
			Matched Filter     & \((3L^{Rx}N_g^{Rx}+1)(L^{Rx}f_{sym})(N_w+N_l+N_b-N_e) + 2N_e\) & \(2L^{Rx}f_{sym}(N_w+N_l+N_b-N_e)+N_e\) & \((L^{Rx})^2N_g^{Rx}f_{sym}\) \\ \hline
			Downsampler        & \(2L^{Rx}f_{sym}(N_w+N_l-N_e)+N_e + (N_w+N_l+N_b)\) & \(2f_{sym}(N_w+N_l+N_b-N_e)+N_e\) & \(0\) \\ \hline
			Symbol Demapper    & \(2f_{sym}(N_w+N_l-N_e)+N_e + \frac{J}{2}(N_w+N_l)\) & \(Jf_{sym}\) & \(0\) \\ \hline
			
		\end{tabular}
\end{table*}

We encode complex block IQ samples in Exponent Box Encoding and retain the floating-point resolution in 32-bit IEEE-754 precision in our model. For simplicity, we select block size to be, \(N_v=L^{TX}f_{sym}\). The symbol mapper generates a \(L^{TX}f_{sym}\)-size of complex block IQ samples that shares common exponent. Pulse shape filter is implemented as Finite Impulse Response (FIR) filter of \(N^{TX}\)-order and requires complex convolution on the upsampled complex block IQ samples.

\subsection{Discrete-time Complex Baseband QAM Receiver}

Due to the channel effect such as fading in practice, the received signals will have larger span in magnitude-phase response. The Common Exponent Encoding applied on sampled complex block IQ samples is limited to selecting window size of minimum phase difference. The Common Exponent Encoding must update its block size at the update rate of gain by the Automatic Gain Control (AGC). Instead, our Exponent Box Encoding could lift the constraint and selects fixed block size, \(N_v=L^{RX}f_{sym}\) in this simulation. We simulate matched filter of \(N^{RX}\)-order.

\section{Simulation Results}

\subsection{Error Vector Magnitude on Complex Block (32-bit) ALU}

Let \(\vec{X}, \vec{\bar{X}} \in \mathbb{C}^{1 \times N}\) be complex-valued row vectors, such that $\vec{X}$ is the reference results in IEEE-754 Encoding and $\vec{\bar{X}}$ is the simulated results in Complex Block Encoding.

The signal quality is measured on the complex block arithmetic results. We truncate the arithmetic results to 32-bit precision to make fair comparison. We use the Root-Mean-Squared (RMS) EVM measurement as described in the following, with \(\parallel \bullet \parallel_2\) as the Euclidean Norm,

\begin{equation} \label{eq7}
\begin{split}
EVM &= \frac{\parallel \vec{X} - \vec{\bar{X}} \parallel_2}{\parallel \vec{X} \parallel_2} * 100
\end{split}
\end{equation}

\begin{figure}
	\includegraphics[width=3.5in]{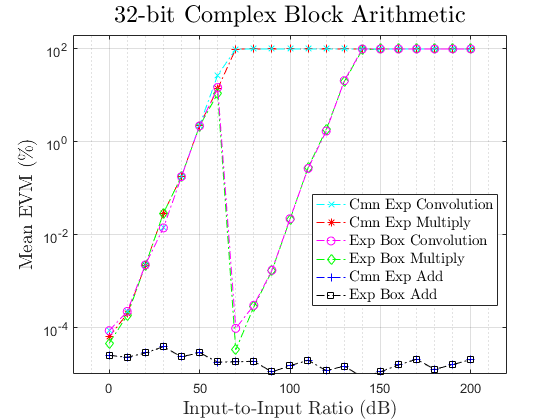}
	\caption{Error vector magnitude of 32-bit complex block arithmetic}
	\label{fig:evmArithmetics}
\end{figure}

Figure \ref{fig:evmArithmetics} shows the EVM of complex block arithmetic in Section \ref{arithmeticSection} on Inputs Ratio \( \in (0,200)\) dB. In complex block addition, the Exponent Box Encoding does not show significant advantage over Common Exponent Encoding because the mantissas addition emphasizes on magnitude over phase. In complex block multiplication and convolution, the Exponent Box Encoding achieves significant reduction in encoding error over Common Exponent Encoding particularly on Inputs Ratio \(\in (70, 140)\) dB where the improvement is between \( (0,99.999)\% \).

\subsection{Error Vector Magnitude on Single-Carrier Transceiver}

\begin{figure}
	\centering
	\includegraphics[width=3.5in]{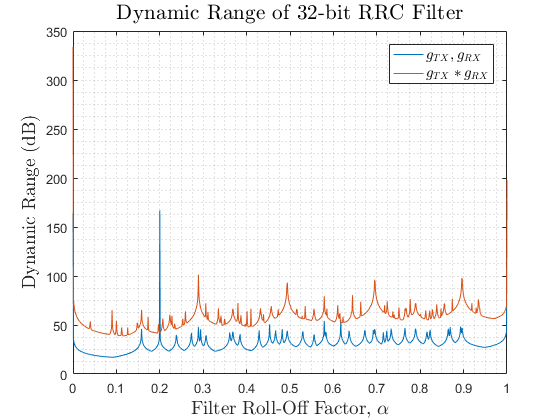}
	\caption{Dynamic range of 32-bit RRC filter impulse response as function of roll-off factor}
	\label{fig:dynamic_range_rrc_filt}
\end{figure}

\begin{figure}
	\centering
	\includegraphics[width=3.5in]{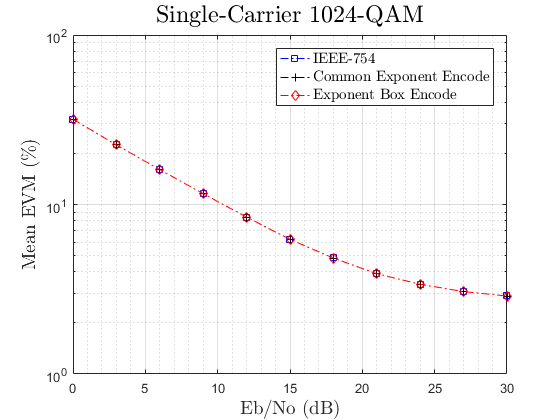}
	\caption{Error vector magnitude between encoding techniques on complex-valued IQ samples}
	\label{fig:EVM_MQAM}
\end{figure}

Figure \ref{fig:dynamic_range_rrc_filt} shows the dynamic range of Root-Raised Cosine (RRC) filter at transmitter and receiver and overall pulse shape response as a function of \(\alpha\). Figure \ref{fig:EVM_MQAM} shows the EVM introduced by Complex Block Encoding under system model defined in Section \ref{systemModel}. The EVM plot is indistinguishable between IEEE-754 Encoding and Complex Block Encoding. The reasons are the selection of RRC Roll-off factor and energy-normalized constellation map.

\section{Conclusion}

Our work has identified the processing overhead of the mantissas and shared exponent in complex block floating-point arithmetic. The common exponent encoding would slightly lower the overhead in complex-valued arithmetic. The box encoding of the shared exponent gives the same quantization errors as common exponent encoding in our case study, which is a 32-bit complex baseband transmitter and receiver. Our work has also quantified memory read/write rates and multiply-accumulate rates in our case study.
Future work could extend a similar approach to representing and processing IQ samples in multi-carrier and multi-antenna communication systems.

\appendices

\section{Derivation of Maximum Exponent Difference Under Common Exponent Encoding Technique}
\label{DerivationMaximumExponentDifferenceAppendix}

Let \(i, j\) be two bounded positive real numbers, representable in floating point precision. Assume that \(i\) has larger magnitude than \(j\), \(|j| < |i|\). Define \(E\{k\}\) as exponent and \(M\{k\}\) as mantissa to \(k\), and \(F(k) = 2^{E\{k\}-1} - 1\) as exponent offset, where \(k=\{i,j\}\). Let \(E\{\Delta\}\) be the difference between two exponents, \((E\{i\} - E\{j\})>0\). 

\begin{equation} \label{eq3}
\begin{split}
j &< i\\
(1.M\{j\}*2^{E\{j\}-F(j)}) &< (1.M\{i\}*2^{E\{i\}-F(i)})\\
(1.M\{j\}*2^{E\{j\}}) &< (1.M\{i\}*2^{E\{i\}})\\
(1.M\{j\}*2^{E\{j\}-E\{i\}+E\{i\}}) &< (1.M\{i\}*2^{E\{i\}})\\
(1.M\{j\}*2^{E\{j\}-E\{i\}}) &< (1.M\{i\})\\
(1.M\{j\}*2^{-E\{\Delta\}}) &< (1.M\{i\})\\
(0.M\{j'\}) &< (1.M\{i\})\\
where \quad M\{j'\} &= \frac{1.M\{j\}}{2^{E\{\Delta\}}} \\
\end{split}
\end{equation}

The mantissa bits in \(M(j')\) are truncated in practice, therefore, \(E\{\Delta\}\) must be less than \(M(j)\). The quantization error is the largest when the \(M(j')\) gets zero when \(M(j)\) is nonzero.

\section{Derivation of Pre / Post Processing Complexity of Complex-valued Convolution}
\label{ComplexConvolutionComplexityAppendix}

Let \(N_{mult}^{mant},N_{add}^{mant},N_{mult}^{exp},N_{add}^{exp}\) be processing complexity of mantissas and exponents determined in Section \ref{arithmeticSection}. 

Among the first and last \(N_1\) terms of \(\vec{Y}\), they are computed by complex inner product of \(i\in\{1,...,N_1\}\) input terms from \(\vec{X_{1}}, \vec{X_{2}}\) and requires \(\frac{(N_1)(N_1+1)}{2}(N_{mult})\) and \(\frac{(N_1-1)(N_1)}{2}(N_{add})\). Among the centering \(N_2-N_1\) terms of \(\vec{Y}\), they are computed by complex inner product of \(N_1\) input terms from \(\vec{X_{1}}, \vec{X_{2}}\) and requires \((N_2-N_1)((N_1)(N_{mult})+(N_1 - 1)(N_{add}))\). 

Overall Multiplication Requirement \((N_{mult})\): 

\begin{equation} \label{eq5}
\begin{split}
&\frac{1}{2}(N_1)(N_1+1) + (N_2-N_1)(N_1) + \frac{1}{2}(N_1-1)(N_1) \\
&=\frac{1}{2}(N_1^2+N_1) + (N_2N_1-N_1^2) + \frac{1}{2}(N_1^2-N_1) \\
&=\frac{1}{2}(2N_1^2) + (N_2N_1-N_1^2) \\
&=N_1^2 + (N_2N_1-N_1^2) \\
&=N_2N_1 \\
\end{split}
\end{equation}

Overall Addition Requirement \((N_{add})\): 

\begin{equation} \label{eq7}
\begin{split}
&\frac{1}{2}(N_1-1)(N_1) + (N_2-N_1)(N_1-1) + \frac{1}{2}(N_1-2)(N_1-1) \\
&=\frac{1}{2}(N_1^2-N_1+N_1^2-3N_1+2) + (N_2-N_1)(N_1-1) \\
&=(N_1^2-2N_1+1) + (N_2-N_1)(N_1-1) \\
&=(N_1-1)(N_1-1) + (N_2-N_1)(N_1-1) \\
&=(N_1-1)(N_1-1 + N_2-N_1) \\
&=(N_1-1)(N_2-1) \\
\end{split}
\end{equation}

Mantissa processing requirement is \((N_{mult}^{mant})(N_2N_1)+(N_{add}^{mant})(N_1-1)(N_2-1)\) and exponent processing requirement is \((N_{mult}^{exp})(N_2N_1)+(N_{add}^{exp})(N_1-1)(N_2-1)\).

\bibliographystyle{IEEEtran}
\bibliography{final_report}

\end{document}